\documentstyle[12pt,epsf,epsfig]{article}
\begin{document}
\title{\large \bf{Radial Oscillations of Hybrid Stars}}
\author{V. K. Gupta \footnote{E--mail :vkg@ducos.ernet.in} ,Vinita Tuli and
 Ashok Goyal \footnote{E--mail :agoyal@ducos.ernet.in}\\ 
         {\em Department of Physics and Astrophysics,}\\
         {\em University of Delhi, Delhi-110 007, India}\\
         {\em Inter-University centre for Astronomy and Astrophysics}\\
         {\em Ganeshkhind, Pune 411007, India. } } 
\renewcommand{\today}{}
\setlength\textwidth{7.5 in}
\setlength\topmargin{-1.cm}
\setlength\textheight{9.5 in}
\addtolength\evensidemargin{-1.cm}
\addtolength\oddsidemargin{-1.cm}
\font\tenrm=cmr10
\def\baselinestretch{1.4}

\maketitle
\begin{abstract}
We study the effect of quark and nuclear matter mixed phase on the radial 
oscillation modes of neutron stars. For this study we have considered recent 
models from two classes of equations of state namely the relativistic mean 
field theoretical models and models based on realistic nucleon-nucleon 
interactions incorporating relativistic corrections and three body nuclear 
interactions.
\end{abstract}
\pagebreak
\begin{section}{Introduction}

Recently there have been important developments in the determination of neutron
star masses in binary systems. Where as the best determination of neutron star 
masses, all in 
the range $1.35\pm0.04$ $M_{\odot}$ (Thorsett 1999), is found for binary radio 
pulsars, limits on the masses of neutron stars from measurements of kilohertz
quasiperiodic oscillation frequencies in low mass X-ray binaries 
lie in the region of 2$M_{\odot}$ (Miller 1998a; 1998b; 1998c; see also 
van-der Klis 1998). In addition several X-ray binary masses 
have also been found to be high viz. Vela X-1 and Cygnus X-2 with masses
$(1.8\pm0.2)$$M_{\odot}$ and $(1.8\pm0.4)$$M_{\odot}$ respectively 
 (van Kerkwijk 1995; Orsoz 1999). These
large masses put severe constraints on the equation of state(EOS) of dense 
matter and only very
stiff EOS's are capable of sustaining such high masses. There are esentially 
two classes of EOS's for dense matter, one class based on relativistic mean 
field
type of models with coupling constants treated as parameters to be fitted to
 observable quantities and the other based on 
realistic 
interactions between constituents obtained by fitting scattering data and using
techniques of many body theory to evaluate energy and pressure. Both classes 
of models however, suffer from lack of sufficient experimental data. This
situation is likely to be ameleorated by our understanding of the physics of
heavy-ion collisions in laboratory experiments in near future. In the core of 
neutron stars where densities rise few times the normal nuclear densities, the
state of matter is essentially unknown and the possibility of a phase
transition to constituent quark matter or to hadronic matter containing 
hyperons and/or meson
condensates has been extensively studied in the litrature (see  for example
Heiselberg 2001 and references therein). All these phase transitions whether first order or second result 
in softening of the EOS and thus aggravating the maximum mass limit. In a 
phase transition to quark matter, the neutron star may have a core of pure 
quark matter with a mantle of nuclear matter surrounding it and the two 
phases co-existing in a first order phase transition. Alternatively one may 
have a so called hybrid star discussed by Glendenning (Glendenning 1992; ibid 
1997; see also Heiselberg 2001), wherein the quark and nuclear matter 
coexist in mixed phase with
continuous pressure and density variation - a situation obtained by applying
Gibb's criteria to a two-component system.\\

  One important aspect of various compact stars is the study of their radial modes of oscillation. Radial oscillations give information about the stability of the stellar structure under consideration, and are thus quite important in distinguishing between various models of the stellar structure. Also as is well known, radial oscillations do not couple to gravitational waves; as a result the equations governing radial oscillations are quite simple and it is relatively easy to solve the eigenvalue problem that leads to the descrete set of 
eigenfrequencies for the system. The eigenfrequencies form a complete set and hence it is possible to describe any arbitrary periodic radial disturbance as a superposition of its various eigenmodes.\\

      For neutron stars, beginning with Chandrasekhar (1964), radial modes 
 have been investigated in the 
literature for more than thirty years and for various nuclear matter 
equations of state. (see for example Chanmugam 1977; Glass and Lindblom 1983; Vath and Chanmugam 1992; Kokkotas 2001)  . In view of the fact that only a 
few EOS viz. those of Wiringa, Fiks and Fabrocini (Wiringa 1988) using A14+UV11
and U14+UV11 interactions and the recent EOS of Akaml, Pandharipande and
Ravenhall (Akaml 1998) incorporating relativistic corrections and three-
nucleon interactions in a new nucleon-nucleon interaction 
model (the Argonne A18 potential) are practically the only viable neutron 
star models capable of giving neutron star masses greater than 1.9 
$M_{\odot}$, we have picked them for the present study. In addition we also consider the
EOS given by Glendenning from the class of relativistic mean field 
theoretical models (Glendenning 1997). These equations invariably allow a 
phase transition in the core. In this paper we study the effect of a mixed 
quark-nuclear matter core on radial oscillations of hybrid stars. 
\end{section}
\begin{section}{Radial Oscillations of a non-rotating Star}
The equations governing the radial oscillations of a non-rotating star, using 
static, spherically symmetric metric
\begin{equation}
d{s^2}=-e^{2\nu}dt^{2}+e^{2\lambda}dr^{2}+r^{2}(d{\theta}^{2}+sin^{2}{\theta}d\phi^{2})
\end{equation}
 were given by Chandrasekhar (Chandrasekhar 1964). The 
structure of the star in hydrostatic equilibrium is determined by the 
Tolman-Openheimer-Volkoff equations
\begin{equation}
\frac{dp}{dr}=\frac{-G(p+\rho)(m+4\pi r^3p)}{r^2(1-\frac{2GM}{r})}
\end{equation}
\begin{equation}
\frac{dm}{dr}=4\pi{r^{2}}\rho
\end{equation}
\begin{equation}
\frac{d\nu}{dr}=\frac{2GM(1+\frac{4\pi{r^3}p}{m})}{r(1-\frac{2GM}{r})}
\end{equation}
where we have put $ c = 1$.
Assuming a radial displacement ${\delta}r$ with harmonic time dependence 
${\delta}r$$\sim$$e^{i{\omega}t}$ and defining 
variables $\xi=\frac{{\delta}r}{r}$ and $\zeta=r^{2}e^{-\nu}\xi$ , 
the equation governing radial adiabatic oscillations is given by
\begin{equation}
F\frac{d^{2}\zeta}{dr^{2}}+G\frac{d\zeta}{dr}+H\zeta=\omega^{2}\zeta
\end{equation}
where
\begin{equation}
F=-\frac{e^{{2\nu}-{2\lambda}}({\gamma}p)}{p+\rho}
\end{equation}
\begin{equation}
G=-\frac{e^{{2\nu}-{2\lambda}}}{p+\rho}\Bigg[{\gamma}p(\lambda+3\nu)+\frac{d({\gamma}p)}{dr}-\frac{2}{r}({\gamma}p)\Bigg]
\end{equation}
\begin{equation}
H=\frac{e^{{2\nu}-{2\lambda}}}{p+\rho}\Bigg[\frac{4}{r}\frac{dp}{dr}+8{\pi}Ge^{2\lambda}p(p+\rho)-\frac{1}{p+\rho}(\frac{dp}{dr})^{2}\Bigg]
\end{equation}
$\lambda$ is related to the metric function through
\begin{equation}
e^{-2\lambda}=(1-\frac{2GM(r)}{r})
\end{equation}
and $\gamma$ is the adiabatic index, related to the speed of sound through
\begin{equation}
\gamma=\frac{p+\rho}{p}\frac{dp}{d\rho}
\end{equation}
Equation(5) is solved under the boundary conditions
\begin{equation}
\zeta(r=0) = 0 \hskip 1.5cm
{\delta}p(r=R) = 0
\end{equation}
where $\delta{p}(r)$ is given by
\begin{equation}
{\delta}p(r)=-\frac{dp}{dr}\frac{e^{\nu}\zeta}{r^{2}} - 
\frac{\gamma p e^{\nu}}{r^2}\frac{d\zeta}{dr}
\end{equation}
Equation (5) with the boundary condition (11) represent a Sturm-Liouville 
eigenvalue problem for $\omega^{2}$ with the well known result that the 
frequency 
spectrum is descrete. For $\omega^2 > 0$, $\omega$ is real and the solution is purely 
oscillatory
whereas for $\omega^2 < 0$, $\omega$ is imaginery resulting in exponentially growing unstable radial
oscillations. Another important consequence is that if the fundamental radial
mode $\omega_0$ is stable, so are the rest of the radial modes. For neutron stars
$\omega_0$ becomes 
imaginery at central densities $\rho_c > \rho_c^{critical}$ for which the star attains it's
maximum mass. For  $\rho_c = \rho_c^{critical}$, the fundamental frequency $\omega_0$ vanishes and becomes unstable for higher densities and the star is no longer stable.  There also
exists another unstable point at the lower end of the central density, namely
there exists a minimum mass for a stable neutron star and the frequency of
the fundamental mode at the minimum mass again goes to zero.
\end{section}
\pagebreak

\begin{section}{Results}
The only information required to obtain the structure of the star and 
eigenvalues of radial oscillation modes is the knowledge of the EOS. For a
given EOS, equations (2)-(5) are solved numerically by standard techniques
under the boundary conditions (11). While numerically integrating the 
equations for each EOS we make sure that the eigenfrequency of the fundamental
mode goes to zero at the maximum mass of the star. We have also checked that
the frequency also vanishes at the minimum stable mass too. For the EOS, as
discussed in the introduction, we use two relativistic mean field theoretic
models taken from Glendenning (Glendenning 1997) and two potential models
incorporating relativistic corrections and three body interactions given by 
Akaml, Pandharipande and Ravenhall (Akaml 1998). Both class of models admit 
of mixed quark-nucleon phase in the core and correspond to matter as below :\\

RFT1H : pure confined hadronic phase for $nB < 0.26 fm^{-3}$, mixed confined 
phase at intermediate densities and pure quark-matter at $nB > 1.17 fm^{-3}$.
Nuclear properties correspond to $K = 240$ MeV and $\frac{m^*}{m} = 0.78$.\\

RFT2H : confined hadronic phase for $nB < 0.26 fm^{-3}$, deconfined phase for 
$nB > 1.0 fm^{-3}$, mixed phase inbetween. $K = 300$ MeV and $\frac{m^*}{m} = 0.78$.\\

The models RFT1 and RFT2 correspond to pure hadronic phase occuring in the 
core of neutron stars without the existence of any quark phase.\\

PMT1H : Incorporating Argonne A18 potential along with three body interaction 
(A18+UIX model of Akaml et al. 1998) corresponds to pure confined hadronic 
phase for $nB < 0.1 fm^{-3}$, mixed phase at intermediate densities and pure 
phase for $nB > 0.49 fm^{-3}$.\\

PMT2H : Incorporating relativistic boost correction $\delta v$ along with 
three body interaction $UIX^*$ with Argonne A18 potential model corresponds to
pure confined hadronic phase for $nB < 0.1 fm^{-3}$, and pure phase for 
$nB > 0.74 fm^{-3}$.\\

The models PMT1 and PMT2 refer to the above two models with pure confined 
hadronic phase.\\

The quark matter in the above EOS models is described by the Bag Model with 
$m_ u = m_ d =0$, $m_ s =150$ MeV, the Bag constant $B^{\frac{1}{4}} = 180$ 
MeV and $\alpha_s = 0$.\\

To illustrate the effect of mixed phase on neutron star parameters, namely 
mass-radius relationship and on the frequency of radial oscillations, we 
have taken one model each from the above two categories considered. since the 
mass fraction contained in the crust of the star is a small fraction of the 
total star mass ($< 2\%$), we have used the earlier results (Baym 1971; Lorenz 1993; Pethick 1995) of the EOS for matter densities $\leq 0.1 fm^{-3}$. In fig
1 we have plotted the M-R curves and find that the effect of the existence of 
mixed quark-nuclear matter phase in the core of neutron stars is to reduce the 
maximum mass. The effect is more pronounced for the relativistic mean field theory model than for the potential model. Whereas for the RFT models also
radius corresponding to maximum mass decreases, for the potential models it increases instead. For smaller mass the two curves RFT2 and RFT2H (PMT2 and PMT2H) come very close to each other; this is to be expected since these points correspond to low central densities by which time the effect of deconfinement becomes negligible. In fig2 we have plotted the frequency of the fundamental mode 
and the next mode as a function of central density for the pure neutron and 
hybrid 
stars in the two categories and find that the frequency exhibits oscillatory
behaviour in the case of a neutron star with a mixed quark-nuclear matter core.
In tables (1a)-(4b), we provide the numerical results for the star parameters and radial mode frequencies of the models discussed in the text. In each table 
we list the central density $\rho_c$, radius R, mass M, red shift 
$Z = (1-\frac{2GM}{R})^{-\frac{1}{2}}-1$ of the 
stellar model and the frequencies of the first two radial models. The model 
above the stability limit is marked by a star.
\end{section}
\begin{section}{Acknowledgements}
We thank J. V. Narlikar for providing hospitality at Inter-University Centre for Astronomy and Astrophysics, Pune. One of us A. G. thanks University Grants Commission, India for partial financial support.
\end{section} 
\pagebreak

\begin{center}
TABLE 1a: RFT1
\vskip 1cm
\begin{tabular}{|c|c|c|c|c|c|}\hline
$\rho_{c}$$\times$$10^{14}$&R&M&Z&$\nu_0$&$nu_1$\\
(gm/cc)&(Km)&($M_{\odot}$)&&(KHz)&(KHz)\\\hline
30.04&10.63&1.546&0.33&0.96&6.22*\\
25.35&10.95&1.550&0.31&0.45&6.32\\
15.22&12.09&1.496&0.257&1.52&6.26\\
10.65&12.78&1.394&0.22&1.89&6.16\\
8.79&13.07&1.314&0.19&2.09&6.14\\
7.00&13.35&1.187&0.17&2.38&6.05\\
5.83&13.48&1.072&0.14&2.65&6.26\\
4.70&13.58&0.868&0.11&2.81&5.67\\
3.61&13.70&0.608&0.07&2.89&4.01\\
3.07&13.86&0.471&0.05&3.04&3.18\\\hline
\end{tabular}\\
\vskip 3cm
TABLE 1b: RFT1H
\vskip 1cm
\begin{tabular}{|c|c|c|c|c|c|}\hline
$\rho_{c}$$\times$$10^{14}$&R&M&Z&$\nu_0$&$\nu_1$\\
(gm/cc)&(Km)&($M_{\odot}$&&(KHz)&(KHz)\\\hline
31.68&10.20&1.45&0.31&0.89&6.30*\\
28.58&10.42&1.45&0.30&0.29&6.12\\
18.50&11.18&1.41&0.26&1.84&6.31\\
15.26&11.56&1.36&0.24&2.04&12.60\\
11.68&12.18&1.24&0.20&2.02&6.65\\
8.97&12.85&1.10&0.16&1.80&6.26\\
6.69&13.38&0.98&0.13&1.90&5.64\\
4.58&13.58&0.84&0.11&2.84&5.49\\
4.15&13.62&0.74&0.09&2.81&4.86\\
3.61&13.69&0.61&0.07&2.89&4.01\\\hline
\end{tabular}\\
\pagebreak

TABLE 2a: RFT2
\vskip 1cm
\begin{tabular}{|c|c|c|c|c|c|}\hline
$\rho_{c}$$\times$$10^{14}$ & R   & M &Z         &$\nu_0$ &$ \nu_1$  \\
gm/cc  &(Km)&($M_{\odot}$)& &(KHz)&(KHz) \\ \hline 
30.82&10.88&1.645&0.346&1.38&5.976*              \\
25.12&11.28&1.659&0.332&0.77&6.072* \\
23.50&11.42&1.661&0.326&0.50&6.076*\\
22.10&11.57&1.662&0.320&0.32&6.084\\
12.07&12.86&1.577&0.253&1.61&5.953\\
8.86&13.35&1.458&0.216&2.01&5.909\\
7.04&13.64&1.319&0.184&2.28&5.741\\
5.84&13.77&1.186&0.159&2.56&5.835\\
4.70&13.85&0.956&0.121&2.66&5.668\\
4.15&13.89&0.807&0.099&2.64&5.066\\
3.62&13.97&0.651&0.077&2.60&4.143\\
3.07&14.17&0.495&0.057&2.51&3.146\\
2.55&14.81&0.344&0.036&2.05&2.394\\\hline
\end{tabular}\\

\vskip 1.5cm

TABLE 2b: RFT2H
\vskip 1.0cm

\begin{tabular}{|c|c|c|c|c|c|}\hline
$\rho_{c}$$\times$$10^{14}$&R&M&Z&$\nu_{0}$&$\nu_1$\\
(gm/cc)&(Km)&($M_{\odot}$)&&(KHz)&(KHz)\\\hline
32.07&10.31&1.430&0.303&0.776&6.27*\\
27.40&10.68&1.434&0.289&0.213&6.13\\
20.88&11.40&1.425&0.261&0.851&5.57\\
17.39&11.68&1.409&0.248&1.426&5.83\\
13.27&12.20&1.346&0.219&1.723&6.16\\
10.92&12.66&1.268&0.193&1.721&6.14\\
8.60&13.22&1.160&0.163&1.641&5.72\\
6.48&13.68&1.041&0.136&1.795&5.09\\
5.26&13.83&0.958&0.122&2.262&5.00\\
4.10&13.89&0.807&0.099&2.643&5.07\\
3.62&13.97&0.651&0.077&2.603&4.14\\
3.07&14.17&0.495&0.056&2.515&3.15\\
2.55&14.81&0.344&0.036&2.054&2.40\\\hline
\end{tabular}\\

\pagebreak
TABLE 3a: PMT1
\vskip 1.0cm
\begin{tabular}{|c|c|c|c|c|c|}\hline
$\rho_{c}$$\times$$10^{14}$&R&M&Z&$\nu_0$&$\nu_{1}$\\
(gm/cc)&(Km)&($M_{\odot}$)&&(KHz)&(KHz)\\\hline
26.47&10.56&2.36&0.72&1.27&7.71*\\
24.33&10.70&2.37&0.70&0.60&7.86\\
20.45&11.02&2.35&0.65&1.53&8.14\\
17.05&11.36&2.30&0.58&2.31&8.38\\
14.08&11.69&2.19&0.48&2.92&8.53\\
11.47&11.98&1.96&0.39&3.40&8.51\\
16.34&12.17&1.61&0.28&3.70&8.21\\
9.17&12.20&1.38&0.23&3.73&7.91\\
8.11&12.21&1.13&0.17&3.67&7.39\\
7.10&12.23&0.87&0.13&3.50&6.52\\
5.21&12.35&0.65&0.09&3.29&5.26\\
4.31&12.80&0.45&0.06&2.82&3.48\\
3.42&13.41&0.35&0.04&2.41&2.65\\\hline
\end{tabular}
\\
\vskip 3cm

TABLE 3b: PMT1H
\vskip 1.0cm
\begin{tabular}{|c|c|c|c|c|c|}\hline
$rho_{c}$$\times$$10^{14}$&R&M&Z&$\nu_0$&$\nu_1$\\
(gm/cc)&(Km)&($M_{\odot}$)&&(KHz)&(KHz)\\\hline
31.40&11.33&1.96&0.43&1.96&5.33*\\
20.59&11.49&1.97&0.43&0.46&6.67\\
13.31&11.89&1.89&0.38&2.51&7.61\\
11.29&12.05&1.80&0.34&3.05&7.70\\
9.07&12.15&1.65&0.29&3.54&7.95\\
8.95&12.18&1.56&0.27&3.71&8.17\\
7.41&12.21&1.33&0.22&3.73&7.81\\
7.30&12.22&1.18&0.18&3.49&7.10\\
6.33&12.23&0.92&0.13&3.54&6.73\\
5.95&12.26&0.82&0.12&3.46&6.31\\
5.39&12.32&0.70&0.10&3.36&5.61\\\hline
\end{tabular}
\\
\pagebreak

TABLE 4a: PMT2
\vskip 1.0cm
\begin{tabular}{|c|c|c|c|c|c|}\hline
$\rho_{c}$$\times$$10^{14}$&R&M&Z&$\nu_{0}$&$\nu_1$\\
(gm/cc)&(Km)&($M_{\odot}$&&(KHz)&(KHz)\\\hline
27.10&10.03&2.188&0.68&0.64&8.49\\
17.49&10.82&2.066&0.52&2.90&8.98\\
13.47&11.22&1.831&0.39&3.55&8.93\\
12.28&11.34&1.711&0.35&3.71&8.81\\
11.14&11.43&1.569&0.30&3.82&8.64\\
10.05&11.51&1.403&0.25&3.88&8.38\\
9.01&11.57&1.220&0.21&3.88&8.01\\
8.01&11.63&1.025&0.16&3.79&7.51\\
5.19&12.17&0.520&0.07&3.11&4.40\\
3.42&14.20&0.269&0.03&1.80&2.22\\\hline
\end{tabular}
\\
\vskip 3cm
TABLE 4b: PMT2H
\vskip 1.0cm
\begin{tabular}{|c|c|c|c|c|c|}\hline
$\rho_{c}$$\times$$10^{14}$&R&M&Z&$\nu_0$&$\nu_1$\\
(gm/cc)&(Km)&($M_{\odot}$)&&(KHz)&(KHz)\\\hline
28.56&10.56&1.908&0.47&1.31&6.76*\\
24.64&10.65&1.911&0.46&0.50&7.25\\
20.04&10.83&1.898&0.44&1.86&7.83\\
17.13&11.00&1.861&0.42&2.48&8.15\\
14.97&11.16&1.801&0.38&2.93&8.30\\
13.24&11.29&1.716&0.35&3.30&8.34\\
11.78&11.40&1.606&0.31&3.62&8.37\\
11.12&11.45&1.542&0.29&3.76&8.43\\
10.49&11.49&1.471&0.27&3.86&8.48\\
9.42&11.55&1.295&0.22&3.89&8.18\\
8.40&11.62&1.104&0.18&3.84&7.74\\
7.42&11.70&0.908&0.14&3.70&7.10\\
6.48&11.84&0.721&0.10&3.48&6.14\\
5.56&12.04&0.589&0.08&3.25&5.05\\
4.65&12.57&0.426&0.05&2.84&3.48\\
3.77&13.68&0.305&0.35&2.14&2.45\\\hline
\end{tabular}
\end{center}
\pagebreak
\pagestyle{empty}
\begin{figure}[ht]
\centerline{
\epsfxsize=9cm \epsfysize=10cm \epsfbox{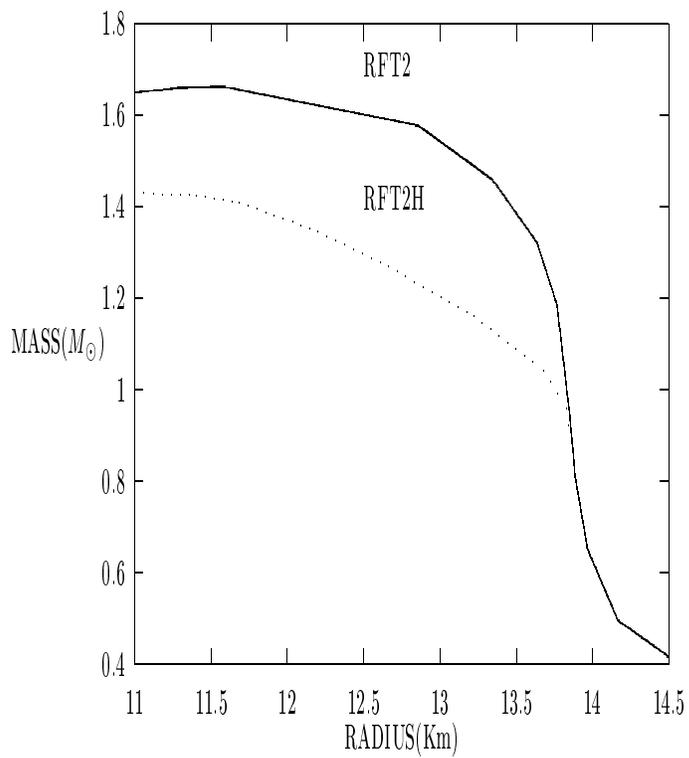}
\hspace*{0.2 cm}
\epsfxsize=9cm \epsfysize=10cm \epsfbox{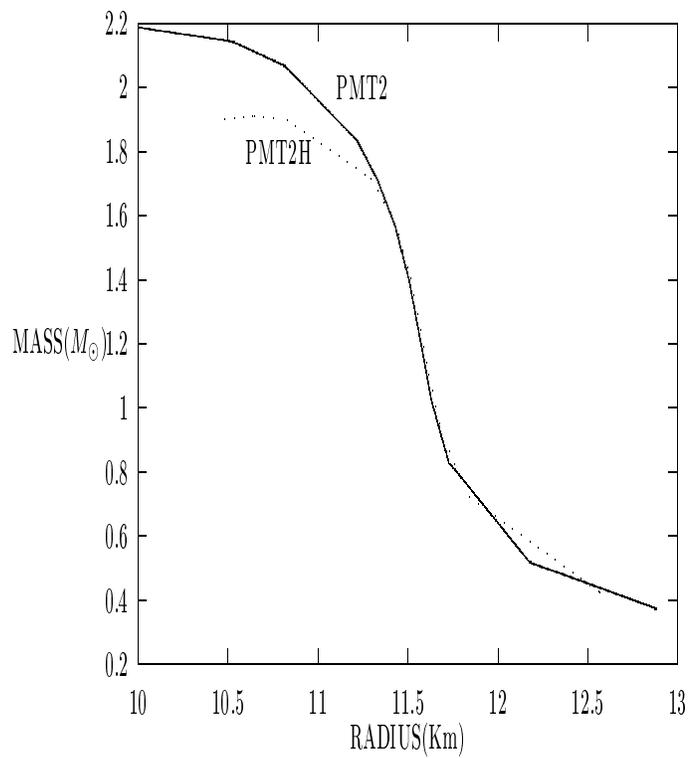}
}
\hskip 1cm
\vskip 1cm
\caption{\em Plot of mass in solar mass unit Vs radius in Km.}
\end{figure}
\pagebreak
\pagestyle{empty}
\begin{figure}[ht]
\centerline{
\epsfxsize=9cm \epsfysize=10cm \epsfbox{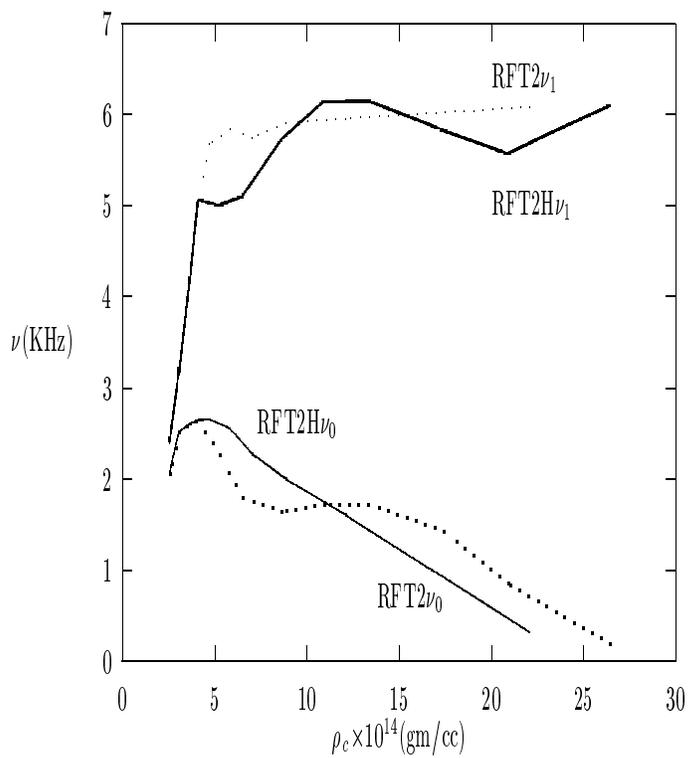}
\hspace*{0.2 cm}
\epsfxsize=9cm \epsfysize=10cm \epsfbox{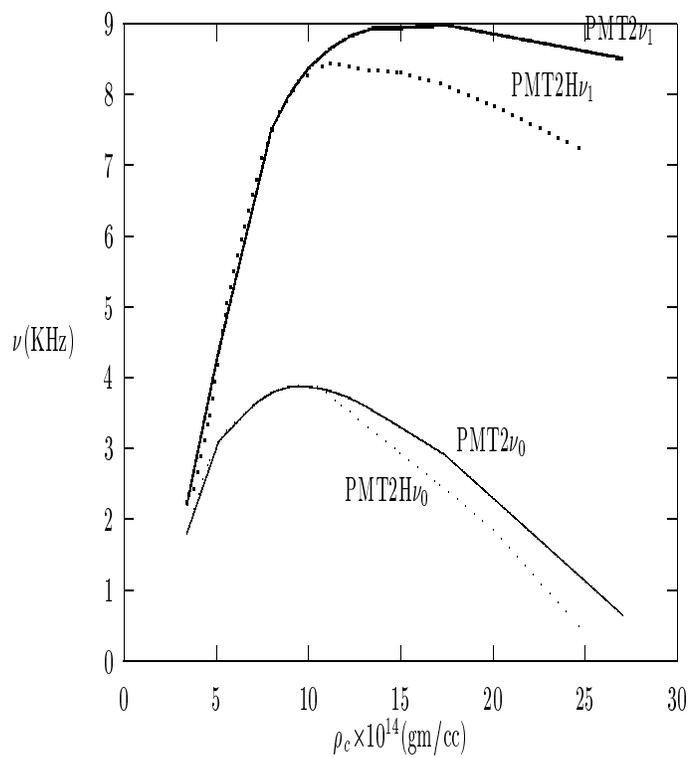}
}
\hskip 1cm
\vskip 1cm
\caption{\em Plot of frequency in KHz Vs central density in gm/cc}
\end{figure}
\pagebreak
   
\end{document}